\documentclass[twoside,11pt]{article}

\usepackage{jmlr2e}

\ShortHeadings{Deep Learning Approaches for Understanding Simple Speech Commands}{Solovyev, Vakhrushev, Radionov, Aliev and Shvets}
\firstpageno{1}

\usepackage[export]{adjustbox}

\begin{document}

\title{Deep Learning Approaches for Understanding Simple Speech Commands}
\author{
      \name Roman A. Solovyev \email turbo@ippm.ru \\
      \addr Institute for Design Problems in Microelectronics of Russian Academy of Sciences \\
      3, Sovetskaya Street, Zelenograd, Moscow, 124365, Russian Federation
      \AND
      \name Maxim Vakhrushev \email vakhrushev.maxim@gmail.com \\
      \addr 2GIS \\
      Novosibirsk, 630048, Russia, 
      \AND
      \name Alexander Radionov \email alex.radionov@gmail.com \\
      \addr 2GIS\\
      Novosibirsk, 630048, Russia,
      \AND
      \name Vladimir Aliev \email vldr.aliev@gmail.com \\
      \addr NGSLab\\
      Moscow, 123022, Russia,
      \AND
      \name Alexey A. Shvets \email shvets@mit.edu \\
      \addr Massachusetts Institute of Technology\\
      Boston, MA 02142, United States of America 
      } 
\maketitle

\begin{abstract}
Automatic classification of sound commands is becoming increasingly important, especially for mobile and embedded devices. Many of these devices contain both cameras and microphones, and companies that develop them would like to use the same technology for both of these classification tasks. One way of achieving this is to represent sound commands as images, and use convolutional neural networks when classifying images as well as sounds. In this paper we consider several approaches to the problem of  sound classification that we applied  in TensorFlow Speech Recognition Challenge organized by Google Brain team on the Kaggle platform. Here we show different representation of sounds (Wave frames, Spectrograms, Mel-Spectrograms, MFCCs) and apply several 1D and 2D convolutional neural networks in order to get the best performance. Our experiments show that we found appropriate sound representation and corresponding convolutional neural networks. As a result we achieved good classification accuracy that allowed us to finish the challenge on 8-th place among 1315 teams.
\end{abstract}

\section{Introduction}

Sound recognition (SR) is the art and science of having machine to identify sounds \citep{jurafsky2014speech}. These sounds could be much beyond than speech and music. Among others it includes such examples as barking dogs, breaking glasses, crying babies and etc. Sound recognition is a key strategic technology that will be embedded in most connected devices offering AI capabilities. For example, every one of us has come across smart-phones with mobile assistants such as Apple Siri, Amazon Alexa or Google Assistant. These applications are dominating and in a way invading human interactions. In addition, there have also been successful uses of voice controlled systems in both medicine and in education for blind and/or handicapped people \citep{okada1998single, mohamed2014educational}. In the nearest future we will see exponential growth of speech recognition embedded devices that will assist to our every day lives. It requires to develop and optimize sound recognition algorithms that need to be fast enough to work in real time and support many different embedded platforms \citep{solovyev2018fpga}.  

Classical approaches in sound recognition system are based on understanding the components of human speech. A phoneme is a contrastive unit in the sound system that helps to distinguish between meanings of words from a set of similar sounds corresponding to it pronounced in one or more ways. For example the word ”speech” has the four phonemes: S P I CH \citep{lee1989speaker, gruhn2011statistical}. To find phonemes, speech signals are slowly timed where their characteristics are stationary over a short period of time. In the feature extraction step, acoustic observations are extracted in frames of typically 25 ms. For the acoustic samples in that frame, a multi-dimensional vector is calculated and on that vector a fast Fourier transformation is performed \citep{lee1989speaker, shrawankar2013techniques}, to transform a function of time, e.g. a signal in this case, into their frequencies.  In the decoding process where calculations is made to find which sequence of words is the most probable match to the feature vectors. For this step, three things has to be present; an acoustic model with a hidden Markov model (HMM) for each unit (phoneme or word), a dictionary containing possible words and their phoneme sequences and a language model with words or word sequences likelihoods \citep{jurafsky2014speech}. Using neural networks as acoustic models for HMM-based speech recognition originally was introduced over 20 years ago \citep{bourlard2012connectionist, mcclelland1986trace}. Much of this original work developed the basic ideas of hybrid DNN-HMM systems which are used in modern, state-of-the-art automatic SR systems \citep{mohamed2012acoustic}. However, until recently, neural networks were not a standard tool in the real time automatic SR systems. Computational constraints and the amount of available training data severely limited the pace at which it was possible to make progress.

Recently, deep learning-based approaches demonstrated performance improvements over conventional machine learning methods for many different applications \citep{lecun2015deep}. The neural networks built with memory capabilities have made speech recognition 99 percent accurate \citep{hinton2012deep, lecun2015deep, goodfellow2016deep}. Neural networks like LSTMs have taken over the field of Natural Language Processing \citep{gers1999learning, graves2005framewise, greff2017lstm}. A person’s speech can also be understood and processed into text by storing the last word of the particular sentence which is fascinating \citep{van2016wavenet, abdel2014convolutional}. Even more the beautiful scientific results now have state-of-the-art applications to the whole process of sound recognition to machine translation \citep{wu2016google, amodei2016deep, bahdanau2014neural}.

In this paper, we consider several approaches based on deep learning to the problem of  sound classification that we applied in TensorFlow Speech Recognition Challenge organized by Google Brain team on Kaggle platform \footnote{\url{https://www.kaggle.com/c/tensorflow-speech-recognition-challenge}}. Here we review 1D convolutional neural networks that uses raw sound files as an input and image-based convolutional neural networks using 2D representation of sounds via Spectrogram, Melgrams and MFCC. We also describe the details of our solution that allowed us to reach pretty good classification accuracy and finish the challenge on 8-th place among 1315 teams.

\begin{figure*}[!t]
\centering
\includegraphics[width=8cm]{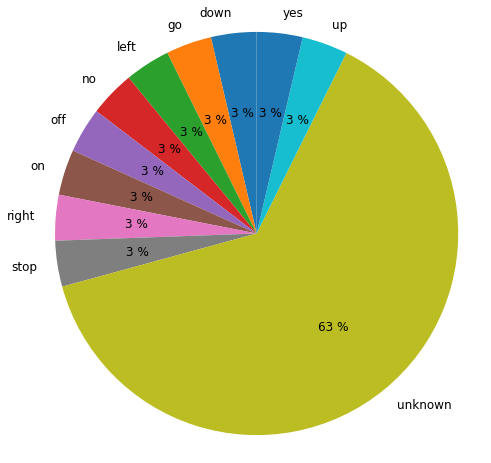}
\caption{Distribution of sound commands in the train data set.}
\label{fig::data_set}
\end{figure*} 

\section{Methods}
\subsection{Data set description and pre-processing}

The training data consists of 60,000 audio files separated by 32 directories with the folder name being the label of the audio clip \citep{warden2018speech}. There are less labels that need to be predicted (12 vs 32 in the train). The labels that should to be identified are \emph{yes}, \emph{no}, \emph{up}, \emph{down}, \emph{left}, \emph{right}, \emph{on}, \emph{off}, \emph{stop}, \emph{go}. Everything else should be considered either \emph{silence} or \emph{unknown} classes. The distribution of classes in the train data set is shown in Fig. \ref{fig::data_set}. Each audio file in the data set is 1-second-long clip of voice commands that is converted into a 16-bit little-endian PCM-encoded WAVE file at a 16000 sample rate. This data set is not completely cleaned up for us. For example, several files in the train data set are not exactly 1 second long. Moreover, there are no \emph{silence} files as such. In these cases we are provided by longer recordings with \emph{background\_noises} that we can split up into 1 second fragments. In addition, one can also mix background noises with word files to get some different augmentations for our sounds during training. The test data set contains an audio folder with 150,000+ files in the format $clip\_000044442.wav$. The task is to predict the correct label. Note that not all of the files are evaluated for the leader-board score.

In the train data set audio files are named so the first element is the subject id of the person who gave the voice command, and the last element indicated repeated commands. Repeated commands are when the subject repeats the same word multiple times. Subject id is not provided for the test data, and you can assume that the majority of commands in the test data were from subjects not seen in train \citep{warden2018speech}. The files contained in the training audio are not uniquely named across labels, but they are unique if you include the label folder. For example, $00f0204f\_nohash\_0.wav$ is found in 14 folders, but that file is a different speech command in each folder. 

One important thing before training models - we need to clean up the data. There are files with pretty low sound volumes. Some of these files are corrupt and only contain noise while others are basically background noise without any spoken word. To remove or correctly re-label these files it helps to sort all files on a dynamic range of the output volume. Then find out if there is a threshold minimum sound level below which all files can be classified as silence. For better performance of the procedure one can manually double check the candidate through listening them and looking at spectrograms.

\begin{figure*}[!t]
\includegraphics[width=\linewidth]{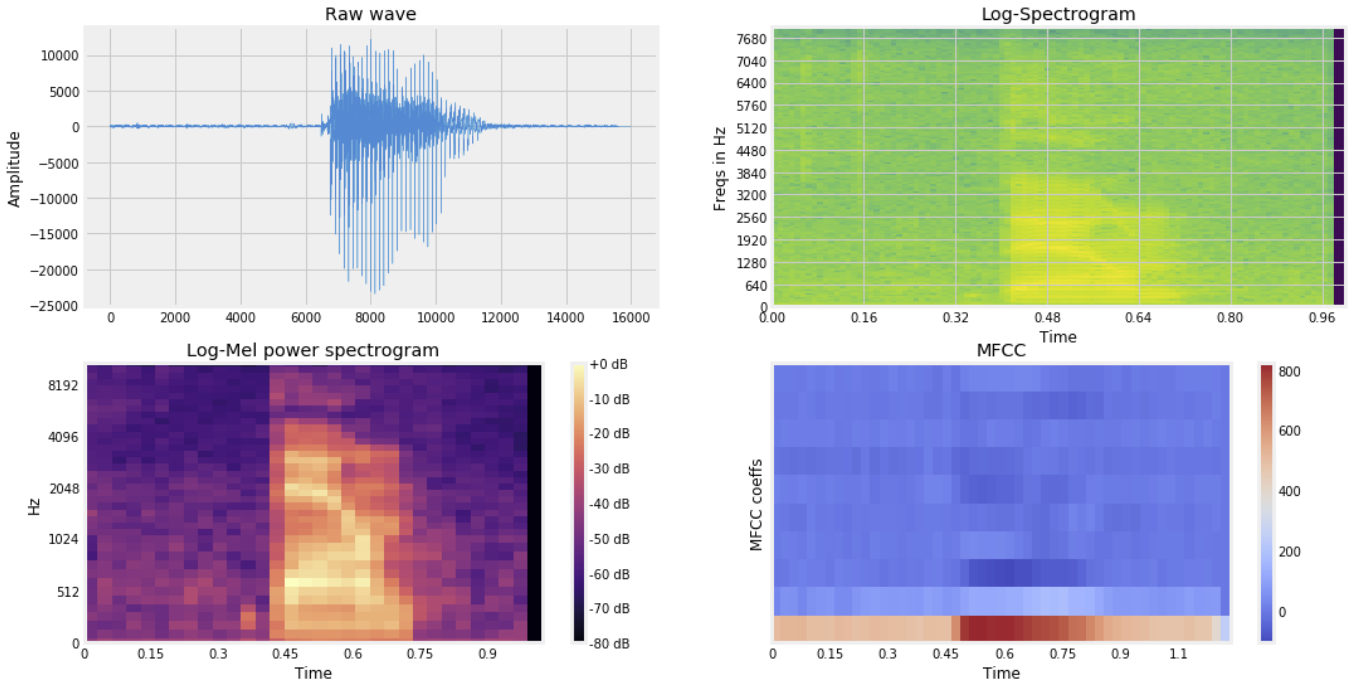}
\caption{Different representations of a sound command that corresponds to the word "Down" ($down/1e4064b8\_nohash\_0.wav$ from the train data set). (Upper-left) The sound signal is a one-dimensional time domain signal; (upper-right) Log-spectrogram using the STFT values with a window size of $25$ ms and step of $10$ ms; (bottom-left) Convert a power spectrogram (amplitude squared) to decibel (dB) units; (bottom-right) Mel-frequency spectral coefficients (MFCC).}
\label{fig::sound_repr}
\end{figure*}

\subsection{Sound representation}

Although deep learning approach eliminates the need for hand-engineered features, we need to choose a representation for our data. The sound signal is a one-dimensional time domain signal. It is difficult to find the rule of how the frequency changes. If we convert the sound signal to frequency domain via Fourier transform (FT), it will show the signal frequency distribution. But at the same time, its time domain information will be missing, making it impossible to see the change of frequency distribution over time. Many joint time-frequency analysis methods have emerged to solve this problem. 

One of the the classic method for joint time-frequency analysis is the short-time Fourier transform (STFT).  STFT is a mathematical transformation associated with FT to determine the frequency and phase of a sine wave in a local region of the time-varying signal. The concept of STFT is to first choose a window function with time-frequency localization. Then assume that the analysis window function $w(t)$ was stationary over a short time, which ensures $f(t)w(t)$ is a stationary signal within different finite time widths. STFT uses fixed window functions, the most commonly used include the Hanning window, the Hamming window, and the Blackman-Haris window \citep{jurafsky2014speech}. The Hamming window, a generalized cosine window, is used in this article. It is usually represented as   
$$
w(t) = a_0+(1-a_0)\cos\left(\frac{2\pi t}{T}\right),\quad 0\le t\le T
$$
where $a_0 = 0.53836$. This function is a member of both the cosine-sum and power-of-sine families. The Hamming window can efficiently reflect the attenuation relationship between energy and time at a certain moment. We show the spectrogram logarithmic the STFT values with a window size of $20$ ms and step of $10$ ms in Fig. \ref{fig::sound_repr}. In addition, we convert a power spectrogram (amplitude squared) to decibel (dB) units. This computes the scaling $10 * log_{10}(S/ref)$ in a numerically stable way (Zeros in the output correspond to positions where $S=ref$).

Spectrograms are usually in the form of a large map. In order to turn the sound features into a suitable size, they often need to be transformed into Mel spectrum via Mel scale filter bank. The Mel spectrum is known that the unit of frequency is Hertz (Hz) and the frequency range of human hearing is $20-20000$ Hz. Human auditory perception does not relate to scale units such as Hz in a linear manner. For example, if we have adapted to a $1000$ Hz sound, then when sound frequency is increased to $2000$ Hz, our ears could only perceive that the frequency may be slightly increased, and we would not realize that the frequency has doubled. The mapping for converting an ordinary frequency scale to Mel-frequency scale is as follows \citep{deng2003speech}:
$$
mel(f) = 2595\,\log_{10}(1+f/700)=1127\,\ln(1+f/700)
$$
Here is a log relationship between Hz and Mel frequency. If the frequency is low, Mel-frequency will change rapidly with Hz otherwise if the frequency is high, Mel-frequency will change slowly. This shows that human ears are sensitive to low frequency sounds and less responsive to high frequency sounds. 

Another  well-known speech  extraction  is  based  on  Mel-frequency Cepstral Coefficients (MFCC). This method is  one  of  the most  popular  feature  extraction  techniques  used  in speech  recognition  based  on  frequency  domain using the  Mel  scale. MFCC is a representation of the real cepstral of a windowed short-time signal derived from the fast FT of that signal. The difference from the real cepstral is that  a  nonlinear  frequency scale is used, which approximates the behaviour of the auditory system. Additionally, these coefficients are robust and reliable to variations according to speakers and recording conditions \citep{dave2013feature}. MFCCs use Mel-scale filter bank where the higher frequency filters have greater bandwidth than the lower frequency filters, but their temporal resolutions are the same \cite{deng2003speech}. The last step is to calculate Discrete Cosine Transformation (DCT) of the outputs from the filter (see Fig. \ref{fig::sound_repr}).

\begin{figure*}[!t]
\centering
\includegraphics[width=\linewidth]{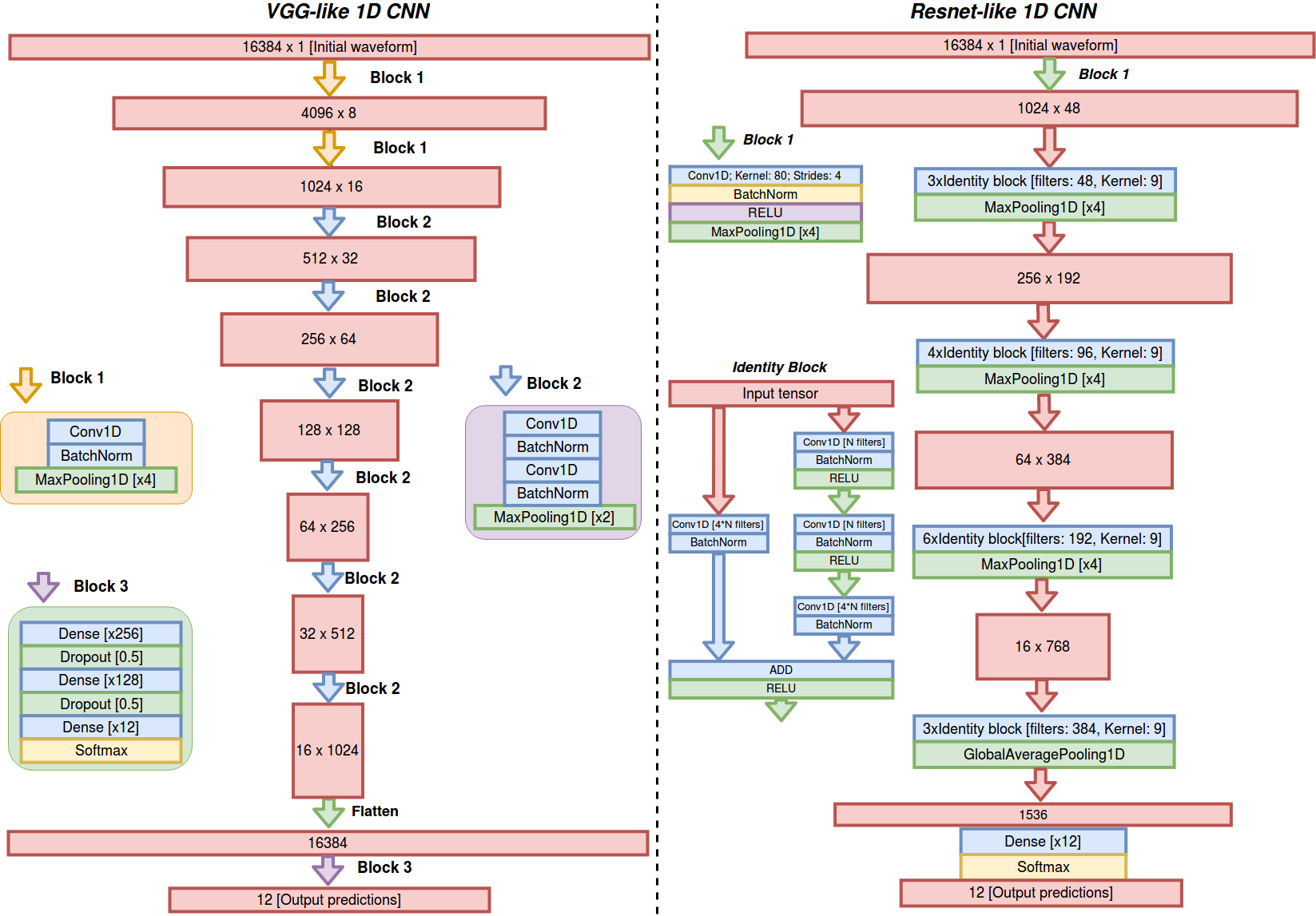}
\caption{Block schemes for VGG-like and ResNet-like 1D convolutional neural networks. Red rectangles represents feature maps while rectangles with other colors correspond to different operations. For VGG-like network the Block 2 contains 2 convolution layers with the same number of output filters. Every convolution operation has a kernel size 9. For this network we use ReLu activation function placed after Batchnorm operation. For the ResNet-like network the identity block is repeated several times depending on size of a feature map.}
\label{fig::cnn1d}
\end{figure*}

In classical, but still state-of-the-art systems, MFCC or similar features are taken as the input to the system instead of spectrograms. However, in end-to-end (often neural-network based) systems, the most common input features are probably raw spectrograms, or mel power spectrograms. For example MFCC decorrelates features, but neural networks deal with correlated features well. As a result, further we will use only raw wave signal for 1D convolution networks, log-spectrograms and mel power spectrograms as an input image for 2D networks. For this work we use scipy\footnote{\url{https://scipy.org/}} and librosa\footnote{\url{https://librosa.github.io/}} python libraries - this code is a standard one for conversion into spectrogram and the software is free to make modifications to suit the needs.

\subsection{1D convolutional neural networks}

Convolutional neural networks (CNN) have demonstrated remarkable success in many Visual Recognition Challenges \citep{shvets2018automatic, rakhlin2018deep, iglovikov2017pediatric}. Such 2D-CNNs learn to recognize the local structure within an image. Inspired by these successes, the hypothesis for this section is that a 1D-CNN will be able to recognize local structure in our time dependant signals. We consider two types of 1D CNN 	
adapted from 2D architectures: VGG \citep{simonyan2014vgg} amd ResNet \citep{he2016resnet} where as an input we use raw signal (see Fig. \ref{fig::cnn1d}). The first type is based on the VGG16 architecture when several convolutional layers alternate with Batchnorm and MaxPooling layers. Then, at the end two blocks of fully connected layers are placed. There are also several differences in comparison to standard VGG16 architecture for 2D images. For example, in the first two layers we use MaxPooling operation that reduce the size of the tensor by 4. It is done to faster reduce dimensionality of the problem. In addition, we have more pooling layers in comparison to 2D case, because the size of input vector is pretty high and we need to reduce it at the end of the network before classification block. As a result the spacial size of the feature map before flatten is equal to 16. In contrast to VGG16 we use single convolution block in the first two layers and then increase it to two consecutive blocks. For activation part we use ReLU activation function placed after Batchnorm operations. As kernel size, we took the value 9, which would be somewhat similar to 3x3 for the case of a two-dimensional convolution. But as far as we understand, it may well work with small values namely, just as 3. Dropout is also applied with the value of 0.5 at the end of the network within fully connected layers to prevent over-fitting.
Similarly to VGG16 we implemented ResNet34/ResNet50 for 1D case. As in the previous case here we also have convolution layers with kernel size equal to 9 for all layers apart from the fist one where the kernel size is equal to 80. In addition, all pooling layers have factor equal to 4. The details of the network are provided in Fig. \ref{fig::cnn1d}.  

\subsection{2D convolutional neural networks}
\begin{figure*}[!t]
\includegraphics[width=\linewidth]{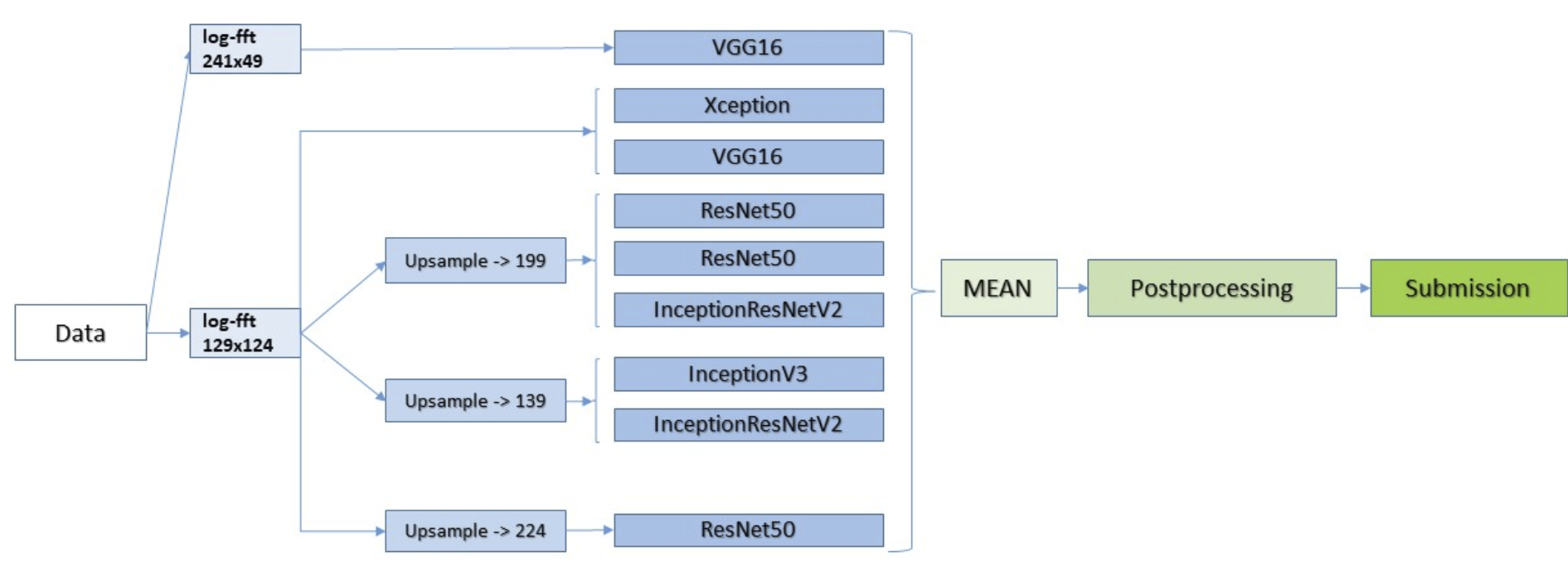}
\caption{2D convolution neural networks that we used in this work and their ensemble.}
\label{fig::ensemble}
\end{figure*}

In 2D case as an input for networks we use log-spectrograms and mel power spectrograms with a standard set of convolutional neural networks available in PyTorch\footnote{\url{https://pytorch.org/}} and Keras\footnote{\url{https://keras.io/}}. We utilize only subset of the networks namely \emph{VGG16}, \emph{VGG19}, \emph{ResNet50}, \emph{InceptionV3}, \emph{Xception}, \emph{InceptionResnetV2} (also shown in Fig.\ref{fig::ensemble}). One very important point is that fine-tuning was forbidden by the competition rules. As a result, here we train all the networks randomly initializing weights. Moreover, we train all of the networks using the same set of hyper-parameters excluding input size that is different.

\section{Results}
We train all models using 4fold cross validation. The data separated between folds by voice id. It means that the voice of one participant did not fall into different folds. Since it was known that in the test set all 12 classes were distributed approximately evenly, we generated each batch with the same number of instances from each class. And for simplicity we chose the size of the batch multiple of 12. 

In speech recognition, data augmentation helps with generalizing models and making them robust against variations in speed, volume, pitch or background noise. In our case augmentations consisted of the following operations: 1) Random change of playback speed rate in the range 0.7 to 1.4; 2) Time shift by a random value in the range -0.1 to +0.1 second; 3) Add random noise from background noise from 0 to 0.05 of the maximum volume of the current waveform. Further, if wave frame was obtained more than the size of the input of a neural network, it was randomly cut off to the desired length. If it turned out less than the desired length, it was supplemented with zeros from the beginning. In 1D cases all audio files are vectors with a length of 16000. First we chose the value 3x4096 = 12288 as the input for the networks and randomly chose such a segment from 16000 as augmentations, but then we realized that most likely it negatively affects the accuracy. As a result the input for neural networks became a vector of length 4x4096 = 16384.  

The quantitative comparison of our models' performance is presented in the Table\ref{table:results}. In this table we present classification accuracy for single network models as well as ensemble of all models for validation, public leader-board and private leader-board datasets. The validation results represent average between 4Folds. One can notice that all the models provide similar performance even those based on 1D convolution neural networks. The ensemble of the models performs better because it reduce variance between models and get rid of outliers. In our case we perform the ensemble as a mean of softmax probabilities. It can be done as a majority vote between the models but its performance is slightly worse. The difference in accuracy between validation and leader-board datastes could be understood if we consider confusion matrix. In such a case we can see that not unknown commands are predicted as unknown, and the predictions are pretty uncertain (0.25-0.4 for unknown, a little less for some other class). We can slighly improve the performance if we play with a threshold for the unknown class. In addition, the test data set contains many files that are purely labeled and said by id's not contained in the train data set. One of the way to improve the performance of the models is to enrich substantially the train dataset by many other examples and somehow improve the quality of signals. 

\begin{table}[t!]
\caption{Classification accuracy (in $\%$) for validation (train), public leader-board (LB) and private LB for several convolution 1D/2D models and their ensemble. Inference time (Time) per image is measured in $ms$ using 1 card gtx 1080ti.}
\label{table:results}
\centering   
\begin{tabular}{|c | c | c | c | c | c |}
\hline
Model & Resolutions & Valid. Acc. & Publ. & Priv. & Time   \\
\hline
VGG-16,\,1D       &1x16384& 93.4 & 86.2 & 86.6 & 0.66 \\
ResNet34,\,1D     &1x16384& 96.4 & 87.6 & 88.0 & 0.82 \\
ResNet50,\,1D     &1x16384& 96.6 & 87.9 & 88.6 & 1.72\\

VGG-16            & 241x49& 94.0 & 87.6 & 88.0 & 1.11\\
Xception          &129x124& 93.9 & 87.4 & 87.9 & 1.71\\
VGG-16            &129x124& 94.4 & 87.7 & 88.1 & 1.22\\
ResNet50          &199x199& 94.2 & 87.5 & 87.9 & 3.19\\
InceptionResnetV2 &199x199& 93.5 & 87.7 & 88.0 & 5.97\\
InceptionV3       &139x139& 94.4 & 87.6 & 88.3 & 2.10\\
InceptionResnetV2 &139x139& 93.7 & 87.3 & 87.8 & 4.62\\
ResNet50          &224x224& 94.7 & 87.8 & 88.4 & 3.55\\
ResNet50,\, mel   &199x199& 93.9 & 87.1 & 87.9 & 3.15\\
Ensemble          &-      & 98.5 & 90.0 & 90.6 &-\\
\hline
\end{tabular}
\end{table}

\section{Conclusions} 
In this study we presented deep learning-based approach for speech command classification in TensorFlow Speech Recognition Challenge organized by Google Brain team on Kaggle platform. We showed different representation of sound command such as wave frames and spectrograms that are used as a input in 1D and 2D convolutional networks. The most popular approaches to this problem are based on fine-tuning of Imagenet networks, but we showed that approaches based on 1D convolutional networks provide very similar performance. This work can be further extended considering different recurrent neural networks architectures, their combinations with convolutional networks and/or siamese networks. Due to the similar performance between several models we can conclude that the data preparation plays a major role for further improvements. Therefore, with the understanding of how to process sound on a machine, one can also work on building their own sound classification systems. The general rule is when it comes to deep learning, the data is the key component. Larger the data, better the accuracy \citep{warden2018speech}. 

\section*{Acknowledgment}
The authors would like to thank Open Data Science\footnote{\url{ods.ai}} and Kaggle\footnote{\url{https://www.kaggle.com/}} communities of great people and colleagues for many valuable discussions and educational help in the growing field of machine/deep learning.

\bibliography{paper}

\begin{thebibliography}{29}
\providecommand{\natexlab}[1]{#1}
\providecommand{\url}[1]{\texttt{#1}}
\expandafter\ifx\csname urlstyle\endcsname\relax
  \providecommand{\doi}[1]{doi: #1}\else
  \providecommand{\doi}{doi: \begingroup \urlstyle{rm}\Url}\fi

\bibitem[Abdel-Hamid et~al.(2014)Abdel-Hamid, Mohamed, Jiang, Deng, Penn, and
  Yu]{abdel2014convolutional}
Ossama Abdel-Hamid, Abdel-rahman Mohamed, Hui Jiang, Li~Deng, Gerald Penn, and
  Dong Yu.
\newblock Convolutional neural networks for speech recognition.
\newblock \emph{IEEE/ACM Transactions on audio, speech, and language
  processing}, 22\penalty0 (10):\penalty0 1533--1545, 2014.

\bibitem[Amodei et~al.(2016)Amodei, Ananthanarayanan, Anubhai, Bai, Battenberg,
  Case, Casper, Catanzaro, Cheng, Chen, et~al.]{amodei2016deep}
Dario Amodei, Sundaram Ananthanarayanan, Rishita Anubhai, Jingliang Bai, Eric
  Battenberg, Carl Case, Jared Casper, Bryan Catanzaro, Qiang Cheng, Guoliang
  Chen, et~al.
\newblock Deep speech 2: End-to-end speech recognition in english and mandarin.
\newblock In \emph{International Conference on Machine Learning}, pages
  173--182, 2016.

\bibitem[Bahdanau et~al.(2014)Bahdanau, Cho, and Bengio]{bahdanau2014neural}
Dzmitry Bahdanau, Kyunghyun Cho, and Yoshua Bengio.
\newblock Neural machine translation by jointly learning to align and
  translate.
\newblock \emph{arXiv preprint arXiv:1409.0473}, 2014.

\bibitem[Bourlard and Morgan(2012)]{bourlard2012connectionist}
Herve~A Bourlard and Nelson Morgan.
\newblock \emph{Connectionist speech recognition: a hybrid approach}, volume
  247.
\newblock Springer Science \& Business Media, 2012.

\bibitem[Dave(2013)]{dave2013feature}
Namrata Dave.
\newblock Feature extraction methods lpc, plp and mfcc in speech recognition.
\newblock \emph{International journal for advance research in engineering and
  technology}, 1\penalty0 (6):\penalty0 1--4, 2013.

\bibitem[Deng and O'Shaughnessy(2003)]{deng2003speech}
Li~Deng and Douglas O'Shaughnessy.
\newblock \emph{Speech processing: a dynamic and optimization-oriented
  approach}.
\newblock CRC Press, 2003.

\bibitem[Gers et~al.(1999)Gers, Schmidhuber, and Cummins]{gers1999learning}
Felix~A Gers, J{\"u}rgen Schmidhuber, and Fred Cummins.
\newblock Learning to forget: Continual prediction with lstm.
\newblock 1999.

\bibitem[Goodfellow et~al.(2016)Goodfellow, Bengio, Courville, and
  Bengio]{goodfellow2016deep}
Ian Goodfellow, Yoshua Bengio, Aaron Courville, and Yoshua Bengio.
\newblock \emph{Deep learning}, volume~1.
\newblock MIT press Cambridge, 2016.

\bibitem[Graves and Schmidhuber(2005)]{graves2005framewise}
Alex Graves and J{\"u}rgen Schmidhuber.
\newblock Framewise phoneme classification with bidirectional lstm and other
  neural network architectures.
\newblock \emph{Neural Networks}, 18\penalty0 (5-6):\penalty0 602--610, 2005.

\bibitem[Greff et~al.(2017)Greff, Srivastava, Koutn{\'\i}k, Steunebrink, and
  Schmidhuber]{greff2017lstm}
Klaus Greff, Rupesh~K Srivastava, Jan Koutn{\'\i}k, Bas~R Steunebrink, and
  J{\"u}rgen Schmidhuber.
\newblock Lstm: A search space odyssey.
\newblock \emph{IEEE transactions on neural networks and learning systems},
  28\penalty0 (10):\penalty0 2222--2232, 2017.

\bibitem[Gruhn et~al.(2011)Gruhn, Minker, and Nakamura]{gruhn2011statistical}
Rainer~E Gruhn, Wolfgang Minker, and Satoshi Nakamura.
\newblock \emph{Statistical pronunciation modeling for non-native speech
  processing}.
\newblock Springer Science \& Business Media, 2011.

\bibitem[He et~al.(2016)He, Zhang, Ren, and Sun]{he2016resnet}
Kaiming He, Xiangyu Zhang, Shaoqing Ren, and Jian Sun.
\newblock Deep residual learning for image recognition.
\newblock In \emph{Proceedings of the IEEE conference on computer vision and
  pattern recognition}, pages 770--778, 2016.

\bibitem[Hinton et~al.(2012)Hinton, Deng, Yu, Dahl, Mohamed, Jaitly, Senior,
  Vanhoucke, Nguyen, Sainath, et~al.]{hinton2012deep}
Geoffrey Hinton, Li~Deng, Dong Yu, George~E Dahl, Abdel-rahman Mohamed, Navdeep
  Jaitly, Andrew Senior, Vincent Vanhoucke, Patrick Nguyen, Tara~N Sainath,
  et~al.
\newblock Deep neural networks for acoustic modeling in speech recognition: The
  shared views of four research groups.
\newblock \emph{IEEE Signal processing magazine}, 29\penalty0 (6):\penalty0
  82--97, 2012.

\bibitem[Iglovikov et~al.(2017)Iglovikov, Rakhlin, Kalinin, and
  Shvets]{iglovikov2017pediatric}
Vladimir Iglovikov, Alexander Rakhlin, Alexandr Kalinin, and Alexey Shvets.
\newblock Pediatric bone age assessment using deep convolutional neural
  networks.
\newblock \emph{arXiv preprint arXiv:1712.05053}, 2017.

\bibitem[Jurafsky and Martin(2014)]{jurafsky2014speech}
Dan Jurafsky and James~H Martin.
\newblock \emph{Speech and language processing}, volume~3.
\newblock Pearson London, 2014.

\bibitem[LeCun et~al.(2015)LeCun, Bengio, and Hinton]{lecun2015deep}
Yann LeCun, Yoshua Bengio, and Geoffrey Hinton.
\newblock Deep learning.
\newblock \emph{nature}, 521\penalty0 (7553):\penalty0 436, 2015.

\bibitem[Lee and Hon(1989)]{lee1989speaker}
K-F Lee and H-W Hon.
\newblock Speaker-independent phone recognition using hidden markov models.
\newblock \emph{IEEE Transactions on Acoustics, Speech, and Signal Processing},
  37\penalty0 (11):\penalty0 1641--1648, 1989.

\bibitem[McClelland and Elman(1986)]{mcclelland1986trace}
James~L McClelland and Jeffrey~L Elman.
\newblock The trace model of speech perception.
\newblock \emph{Cognitive psychology}, 18\penalty0 (1):\penalty0 1--86, 1986.

\bibitem[Mohamed et~al.(2012)Mohamed, Dahl, Hinton,
  et~al.]{mohamed2012acoustic}
Abdel-rahman Mohamed, George~E Dahl, Geoffrey Hinton, et~al.
\newblock Acoustic modeling using deep belief networks.
\newblock \emph{IEEE Trans. Audio, Speech \& Language Processing}, 20\penalty0
  (1):\penalty0 14--22, 2012.

\bibitem[Mohamed et~al.(2014)Mohamed, Hassanin, and
  Othman]{mohamed2014educational}
Samir A~Elsagheer Mohamed, Allam~Shehata Hassanin, and Mohamed Tahar~Ben
  Othman.
\newblock Educational system for the holy quran and its sciences for blind and
  handicapped people based on google speech api.
\newblock \emph{Journal of Software Engineering and Applications}, 7\penalty0
  (03):\penalty0 150, 2014.

\bibitem[Okada et~al.(1998)Okada, Tanaba, Yamauchi, and Sato]{okada1998single}
Shinichiro Okada, Yoshiaki Tanaba, Hideyuki Yamauchi, and Shoichi Sato.
\newblock Single-surgeon thoracoscopic surgery with a voice-controlled robot.
\newblock \emph{The Lancet}, 351\penalty0 (9111):\penalty0 1249, 1998.

\bibitem[Rakhlin et~al.(2018)Rakhlin, Shvets, Iglovikov, and
  Kalinin]{rakhlin2018deep}
Alexander Rakhlin, Alexey Shvets, Vladimir Iglovikov, and Alexandr~A Kalinin.
\newblock Deep convolutional neural networks for breast cancer histology image
  analysis.
\newblock \emph{arXiv preprint arXiv:1802.00752}, 2018.

\bibitem[Shrawankar and Thakare(2013)]{shrawankar2013techniques}
Urmila Shrawankar and Vilas~M Thakare.
\newblock Techniques for feature extraction in speech recognition system: A
  comparative study.
\newblock \emph{arXiv preprint arXiv:1305.1145}, 2013.

\bibitem[Shvets et~al.(2018)Shvets, Rakhlin, Kalinin, and
  Iglovikov]{shvets2018automatic}
Alexey Shvets, Alexander Rakhlin, Alexandr~A Kalinin, and Vladimir Iglovikov.
\newblock Automatic instrument segmentation in robot-assisted surgery using
  deep learning.
\newblock \emph{arXiv preprint arXiv:1803.01207}, 2018.

\bibitem[Simonyan and Zisserman(2014)]{simonyan2014vgg}
Karen Simonyan and Andrew Zisserman.
\newblock Very deep convolutional networks for large-scale image recognition.
\newblock \emph{arXiv preprint arXiv:1409.1556}, 2014.

\bibitem[Solovyev et~al.(2018)Solovyev, Kalinin, Kustov, Telpukhov, and
  Ruhlov]{solovyev2018fpga}
Roman~A Solovyev, Alexandr~A Kalinin, Alexander~G Kustov, Dmitry~V Telpukhov,
  and Vladimir~S Ruhlov.
\newblock Fpga implementation of convolutional neural networks with fixed-point
  calculations.
\newblock \emph{arXiv preprint arXiv:1808.09945}, 2018.

\bibitem[Van Den~Oord et~al.(2016)Van Den~Oord, Dieleman, Zen, Simonyan,
  Vinyals, Graves, Kalchbrenner, Senior, and Kavukcuoglu]{van2016wavenet}
A{\"a}ron Van Den~Oord, Sander Dieleman, Heiga Zen, Karen Simonyan, Oriol
  Vinyals, Alex Graves, Nal Kalchbrenner, Andrew~W Senior, and Koray
  Kavukcuoglu.
\newblock Wavenet: A generative model for raw audio.
\newblock In \emph{SSW}, page 125, 2016.

\bibitem[Warden(2018)]{warden2018speech}
Pete Warden.
\newblock Speech commands: A dataset for limited-vocabulary speech recognition.
\newblock \emph{arXiv preprint arXiv:1804.03209}, 2018.

\bibitem[Wu et~al.(2016)Wu, Schuster, Chen, Le, Norouzi, Macherey, Krikun, Cao,
  Gao, Macherey, et~al.]{wu2016google}
Yonghui Wu, Mike Schuster, Zhifeng Chen, Quoc~V Le, Mohammad Norouzi, Wolfgang
  Macherey, Maxim Krikun, Yuan Cao, Qin Gao, Klaus Macherey, et~al.
\newblock Google's neural machine translation system: Bridging the gap between
  human and machine translation.
\newblock \emph{arXiv preprint arXiv:1609.08144}, 2016.

\end{thebibliography}

\end{document}